\documentclass[12pt]{article}
\usepackage{epsf}
\usepackage{epsfig}
\usepackage{wrapfig}
\usepackage{latexsym}
\usepackage{amsmath}
\usepackage{latexsym}
\usepackage{amsfonts}
\usepackage{amssymb}
\usepackage{amsbsy}
\usepackage[bookmarksnumbered,plainpages]{hyperref}
\begin{document}
\def\bfgr#1{\pmb{#1}}
\def\p{\partial}
\def\dfrac#1#2{{\displaystyle\frac{#1}{#2}}}
\def\stTD#1#2{\hbox to 0em{\mathsurround=0em $\stackrel{#1}{\makebox[0pt]{} #2}$\hss} \phantom{#2}}\def\stscript#1#2{\hbox to 0em{\mathsurround=0em ${\scriptstyle\stackrel{#1}{\makebox[0pt]{} #2}}$\hss} \phantom{#2}}\def\stscriptscript#1#2{\hbox to 0em{\mathsurround=0em ${\scriptscriptstyle\stackrel{#1}{\makebox[0pt]{} #2}}$\hss} \phantom{#2}}\def\st#1#2{\mathchoice{\stTD{#1}{#2}}{\stTD{#1}{#2}}{\stscript{#1}{#2}}{\stscriptscript{#1}{#2}}}
\def\comb#1#2#3{{\mathsurround 0pt\hbox to 0pt {\hspace*{#3}\raisebox{#2}{${#1}$}\hss}}}
\def\combs#1#2#3{{\mathsurround 0pt\hbox to 0pt {\hspace*{#3}\raisebox{#2}{${\scriptstyle #1}$}\hss}}}
\def\combss#1#2#3{{\mathsurround 0pt\hbox to 0pt {\hspace*{#3}\raisebox{#2}{${\scriptscriptstyle #1}$}\hss}}}
\def\e#1{\mathrm{e}^{#1}}
\def\ii{\mathrm{i}}
\def\veps{\varepsilon}
\def\metr{\mathfrak{m}}
\def\cEp{\mathchoice{\combs{\boldsymbol{-}}{0.2ex}{-0.1ex}\mathcal{E}}{\combs{\boldsymbol{-}}{0.2ex}{-0.1ex}\mathcal{E}}{\combss{\boldsymbol{-}}{0.12ex}{-0.18ex}\mathcal{E}}{}{}}
\def\cEt{\mathchoice{\combs{\boldsymbol{\sim}}{0.2ex}{-0.1ex}\mathcal{E}}{\combs{\boldsymbol{\sim}}{0.2ex}{-0.1ex}\mathcal{E}}{\combss{\boldsymbol{\sim}}{0.12ex}{-0.12ex}\mathcal{E}}{}{}}
\def\cP{\mathcal{P}}
\def\bcP{\boldsymbol{\mathcal{P}}}
\def\cE{\mathcal{E}}
\def\Cn{\bar{\C}}
\def\Cm{\mathchoice{\combs{m}{0.45ex}{0.07em}\mathrm{C}}{\combs{m}{0.45ex}{0.07em}\mathrm{C}}{\combss{m}{0.25ex}{0.04em}\mathrm{C}}{}{}}
\def\Ce{\mathchoice{\combs{e}{0.45ex}{0.22em}\mathrm{C}}{\combs{e}{0.45ex}{0.22em}\mathrm{C}}{\combss{e}{0.3ex}{0.15em}\mathrm{C}}{}{}}
\def\C{\mathchoice{\comb{\boldsymbol{\cdot}}{0.22ex}{0.23em}\mathrm{C}}{\comb{\boldsymbol{\cdot}}{0.22ex}{0.23em}\mathrm{C}}{\combs{\boldsymbol{\cdot}}{0.14ex}{0.15em}\mathrm{C}}{}{}}
\def\OOO#1#2{\mathcal{O}\{#1\}_{#2}}
\def\ooo#1#2{o\{#1\}_{#2}}
\def\AMT{M}
\def\EMT{T}
\def\EMV{\mathbb{P}}
\def\Energy{\mathbb{E}}
\def\Vols{V}
\def\dVols{{\rm d}\hspace{-0.2ex}\Vols}
\def\average#1#2#3#4{\mathchoice{\left\langle\mspace{-#4 mu}\left| #1 \right|\mspace{-#4 mu}\right\rangle^{#2}_{#3}}{\left\langle\mspace{-#4 mu}\left| #1 \right|\mspace{-#4 mu}\right\rangle^{#2}_{#3}}{\left\langle\mspace{-#4 mu}\left| #1 \right|\mspace{-#4 mu}\right\rangle^{#2}_{#3}}{}{}}
\def\Vol{\mathchoice{\combs{\square}{0.2ex}{0.2ex}\mathrm{V}}{\combs{\square}{0.2ex}{0.2ex}\mathrm{V}}{\combss{\square}{0.12ex}{0.095ex}\mathrm{V}}{}{}}
\def\dVol{\mathchoice{\mathrm{d}\hspace{-0.3ex}\Vol}{\mathrm{d}\hspace{-0.3ex}\Vol}{\mathrm{d}\hspace{-0.3ex}\Vol}{}{}}
\def\AMD{\mathcal{M}}
\def\bAMD{\boldsymbol{\mathcal{M}}}
\def\AMB{\mathbb{M}}
\def\bAMB{\pmb{\mathbb{M}}}
\def\Surc{\mathcal{S}}
\def\dSurc{\mathrm{d}\hspace{-0.3ex}\Surc}
\def\dSurcb{\mathbf{d}\hspace{-0.3ex}\pmb{\Surc}}
\def\EMT{T}
\def\AMT{M}
\def\xz{t}
\def\xp#1{\comb{\cdot}{-0.9ex}{0.3ex}{#1}}
\def\xpp#1{\comb{\cdot\cdot}{-0.9ex}{0.1ex}{#1}}
\def\xc{\mathchoice{\comb{\boldsymbol{\cdot}}{-0.1ex}{-0.05ex}x}{\comb{\boldsymbol{\cdot}}{-0.1ex}{-0.05ex}x}{\combs{\boldsymbol{\cdot}}{-0.05ex}{-0.05ex}x}{}{}}
\def\Ce{\mathchoice{\combs{e}{0.45ex}{0.225em}C}{\combs{e}{0.45ex}{0.225em}C}{\combss{e}{0.3ex}{0.15em}C}{}{}}
\def\Cm{\mathchoice{\combs{m}{0.45ex}{0.08em}C}{\combs{m}{0.45ex}{0.08em}C}{\combss{m}{0.3ex}{0.02em}C}{}{}}
\def\De{\mathchoice{\combs{e}{0.45ex}{0.75ex}D}{\combs{e}{0.45ex}{0.75ex}D}{\combss{e}{0.3ex}{0.5ex}D}{}{}}
\def\bDe{\mathchoice{\combs{\mathbf{e}}{0.45ex}{0.65ex}\mathbf{D}}{\combs{\mathbf{e}}{0.45ex}{0.65ex}\mathbf{D}}{\combss{\mathbf{e}}{0.3ex}{0.4ex}\mathbf{D}}{}{}}
\def\Dm{\mathchoice{\combs{m}{0.45ex}{0.5ex}D}{\combs{m}{0.45ex}{0.5ex}D}{\combss{m}{0.3ex}{0.3ex}D}{}{}}
\def\bDm{\mathchoice{\combs{\mathbf{m}}{0.45ex}{0.4ex}\mathbf{D}}{\combs{\mathbf{m}}{0.45ex}{0.4ex}\mathbf{D}}{\combss{\mathbf{m}}{0.3ex}{0.2ex}\mathbf{D}}{}{}}
\def\bD{\mathchoice{\comb{\boldsymbol{\cdot}}{0.22ex}{0.3em}\mathbf{D}}{\comb{\boldsymbol{\cdot}}{0.22ex}{0.3em}\mathbf{D}}{\combs{\boldsymbol{\cdot}}{0.14ex}{0.2em}\mathbf{D}}{}{}}
\def\metr{\mathfrak{m}}
\def\Fem{F}
\def\bFem{\mathbf{F}}
\def\fem{G}
\def\bfem{\mathbf{G}}
\def\bcdot{\mathchoice{\mathbin{\boldsymbol{\comb{\cdot}{0.22ex}{0.57ex}\Diamond}}}{\mathbin{\boldsymbol{\comb{\cdot}{0.22ex}{0.57ex}\Diamond}}}{\mathbin{\boldsymbol{\combs{\cdot}{0.13ex}{0.35ex}\Diamond}}}{}{}}
\def\bwedge{\mathchoice{\mathbin{\combs{\boldsymbol{\backslash}}{0.2ex}{0.1ex}\boldsymbol{\wedge}}}{\mathbin{\combs{\boldsymbol{\backslash}}{0.2ex}{0.1ex}\boldsymbol{\wedge}}}{\mathbin{\combss{\boldsymbol{\backslash}}{0.1ex}{0ex}\boldsymbol{\wedge}}}{}{}}
\def\Eem{E}
\def\bEem{\mathbf{E}}
\def\Hem{H}
\def\bHem{\mathbf{H}}
\def\Dem{D}
\def\bDem{\mathbf{D}}
\def\Bem{B}
\def\bBem{\mathbf{B}}
\def\baab{\mathbf{b}}
\def\bbaab{{\mathsurround 0pt\mbox{${\bf b}$\hspace*{-1.1ex}${\bf b}$}}}
\def\unitc{\mathbf{1}}
\def\him{\imath}
\def\bhim{\boldsymbol{\imath}}
\def\hconj#1{\mathchoice{{{}^{\boldsymbol{*}}\mspace{-2mu}#1}}{{{}^{\boldsymbol{*}}\mspace{-2mu}#1}}{{{}^{\boldsymbol{*}}\mspace{-4mu}#1}}{}{}}
\def\bDB{\mathbf{Y}}
\def\bEH{\mathbf{Z}}
\def\prodf#1#2{\mathchoice{\pmb{\Bigl\langle} #1\, \pmb{|}\, #2 \pmb{\Bigr\rangle}}{\pmb{\bigl\langle} #1\, \pmb{|}\, #2 \pmb{\bigr\rangle}}{\pmb{\langle} #1\, \pmb{|}\, #2 \pmb{\rangle}}{}{}}
\def\bShST{\mathchoice{\combs{\rightarrow}{0.2ex}{0.5ex} \boldsymbol{\mathcal{J}}}{\combs{\rightarrow}{0.2ex}{0.5ex} \boldsymbol{\mathcal{J}}}{\combss{\rightarrow}{0.12ex}{0.2ex} \boldsymbol{\mathcal{J}}}{}{}}
\def\bRotS{\mathchoice{\comb{\circ}{0.05ex}{0.78ex} \boldsymbol{\mathcal{J}}}{\comb{\circ}{0.05ex}{0.78ex} \boldsymbol{\mathcal{J}}}{\combs{\circ}{0.05ex}{0.55ex} \boldsymbol{\mathcal{J}}}{}{}}
\def\Fourier#1{{}_{\mathtt{f}}\!\mspace{-1.5mu}#1}
\def\garmcb#1#2#3#4{\mathchoice{{}_#3\!\combs{\combs{\|}{0ex}{0.27ex}{\approx}}{0.4ex}{0.41ex}{\mathbf{C}^{#4}_{#1#2}}}{{}_#3\!\combs{\combs{\|}{0ex}{0.27ex}{\approx}}{0.4ex}{0.41ex}{\mathbf{C}^{#4}_{#1#2}}}{{}_#3\!\combss{\combss{\|}{0.04ex}{0.17ex}{\approx}}{0.2ex}{0.25ex}{\mathbf{C}^{#4}_{#1#2}}}{}{}}
\def\garmsb#1#2#3#4{\mathchoice{{}_#2\!\combs{\comb{\circ}{-0.2ex}{0.06ex}{\approx}}{0.4ex}{0.41ex}{\mathbf{C}^{#3#4}_{#1}}}{{}_#2\!\combs{\comb{\circ}{-0.2ex}{0.06ex}{\approx}}{0.4ex}{0.41ex}{\mathbf{C}^{#3#4}_{#1}}}{{}_#2\!\combss{\combs{\circ}{-0.05ex}{0.13ex}{\approx}}{0.2ex}{0.25ex}{\mathbf{C}^{#3#4}_{#1}}}{}{}}
\def\xcm{\mathchoice{\combs{\circ}{0.13ex}{0.22ex}{x}}{\combs{\circ}{0.13ex}{0.22ex}{x}}{\combss{\circ}{0.06ex}{0.105ex}{x}}{}{}}
\def\xpi#1{{\underline{#1}}}
\def\mass{\mathtt{m}}
\def\spin{\mathtt{s}}
\def\bspin{\mathbf{s}}
\def\partsol#1{\mathchoice{\combs{\circ}{1.65ex}{0.5ex}{#1}}{\combs{\circ}{1.65ex}{0.5ex}{#1}}{\combss{\circ}{1.2ex}{0.3ex}{#1}}{}{}}
\def\df{\mathrm{d}}
\def\dfs{\mathchoice{\combs{|}{0.45ex}{0.1em}\mathrm{d}}{\combs{|}{0.45ex}{0.1em}\mathrm{d}}{\combss{|}{0.25ex}{0.05em}\mathrm{d}}{}{}}
\begin{center}
{\bfseries MASS, SPIN, CHARGE, AND MAGNETIC MOMENT\\ FOR ELECTROMAGNETIC PARTICLE}
\vskip 5mm
\underline{A.A. Chernitskii}
\vskip 5mm
{\small
{\it
A. Friedmann Laboratory for Theoretical Physics, St.-Petersburg
}\\
{\it
%E-mail: AAChernitskii@eltech.ru , Alexander@cher.etu.spb.ru
E-mail: alexander@cher.etu.spb.ru
}}
\end{center}
\vskip 5mm
\begin{abstract}
Electromagnetic particle is considered as appropriate particle solution of nonlinear electrodynamics.
Mass, spin, charge, and dipole moment for the electromagnetic particle are defined.
Classical motion equations for massive charged particle
with spin and dipole moment are obtained from integral conservation laws for the field.
\end{abstract}

We can consider some elementary particle as electromagnetic if it has the specific electromagnetic
properties such that charge and dipole (in particular, magnetic) moment. Also the elementary particle has mass and spin.
These are four basic characteristics for electromagnetic particle.
But charge can be compensated in complicated particles. Thus we can consider also
neutral electromagnetic particles.

Here a model of electromagnetic particle in the framework of Born-Infeld type nonlinear electrodynamics is considered
\cite{Chernitskii1999,Chernitskii2004a}.

General system of electrodynamic equations in inertial coordinates for spatial regions without singularities
has the following form:
\begin{eqnarray}
\nonumber
\frac{\p\bBem}{\p x^{0}}  {}+{}  \mathrm{curl}\bEem  = 0
\;\;,&\qquad
\mathrm{div} \bBem  = 0
\;\;,\\[1ex]
\frac{\p\bDem}{\p x^{0}}  {}-{}  \mathrm{curl}\bHem  = 0
\;\;,&\qquad
\mathrm{div} \bDem  = 0
\;\;,
\label{Eq:Maxwell}
\end{eqnarray}
where $\bDem$ and $\bBem$ are electric and magnetic inductions, $\bEem$ and $\bHem$ are electric and magnetic
 intensities.
%%%%%%%%%%%%%%%%%%%%%%%%%%%%%%%%%%%%%%%%%%%%%%%%%%%%%%%%%%%%

System (\ref{Eq:Maxwell}) must be supplemented with algebraic relations for connection between the fields
$\bDem$, $\bBem$, $\bEem$, $\bHem$.
It is convenient to use the electromagnetic inductions as dependent variables \cite{Chernitskii1999}.
In this case
\begin{equation}
\label{46058534}
\Eem_{i} =4\pi\,\frac{\p \cE}{\p \Dem^{i}}
\;\;,\quad
\Hem_{i} =4\pi\,\frac{\p \cE}{\p \Bem^{i}}
\;\;,
\end{equation}
where $\cE = \cE (\bDem,\bBem)$ is an energy density associate with equation system (\ref{Eq:Maxwell}).
The Latin indices take the values $1,2,3$.

We can write algebraic relation (\ref{46058534}) in the following general form
\begin{equation}
\label{47109986}
\bEem = \bEem (\bDem,\bBem)
\;\;,\quad
\bHem = \bHem (\bDem,\bBem)
\;\;.
\end{equation}
For the case of small field these relations must be asymptotically linear:
\begin{equation}
\label{65158987}
\bEem = \bDem
\;\;,\quad
\bHem = \bBem
\;\;.
\end{equation}

In addition, the hypercomplex form for electrodynamics is convenient to use
\cite{Chernitskii2005a}. In this representation there is one hypercomplex equation
(instead of system (\ref{Eq:Maxwell})) for two quasibivectors
\begin{equation}
\label{36281628}
\bDB \doteqdot \bDem + \bhim\,\bBem
\;\;,\quad
\bEH \doteqdot \bEem + \bhim\,\bHem
\;\;,
\end{equation}
where $\bhim$ is hyperimaginary unit or unit pseudoscalar, $\bhim^2=-1$.
Here $\bhim$ is used as customary imaginary unit (for more details, see \cite{Chernitskii2005a}).

Let us write two differential laws associate with equation system (\ref{Eq:Maxwell}):
\begin{equation}
\label{68098837}
\frac{\p \EMT^{\mu\nu}}{\p x^{\nu}} = 0
\;\;,\quad
\frac{\p \AMT^{\mu\nu\rho}}{\p x^{\rho}} = 0
\;\;.
\end{equation}
The Greek indices take the values $0,1,2,3$.

The components for symmetrical energy-momentum density tensor are
\begin{eqnarray}
\nonumber
&&\EMT^{00} \doteqdot {\cE}
\;\;,\quad
\EMT^{0i} = \EMT^{i0}  \doteqdot {\cP}^i\;\;,\quad
{\bcP} =   \dfrac{1}{4\pi}\left(\bDem\times \bBem\right)
\;\;,
\\
&&\EMT^{ij} = \dfrac{1}{4\pi}\left[
{\metr}^{ij}\left(\Dem^l\,\Eem_l {}+{} \Bem^l\,\Hem_l \right)
{}-{}
\left(\Dem^i\,\Eem^j {}+{} \Bem^i\,\Hem^j\right)\right]
{}-{} \metr^{ij}\,\cE
\;\;,
\end{eqnarray}
where ${\bcP}$ is momentum density,
${\metr}^{\mu\nu}$ are components of metric tensor for inertial coordinate system (${\metr}^{0i}=0$, $|{\metr}^{00}|=1$).
The function of energy density $\cE (\bDem,\bBem)$ defines a specific electrodynamics model
according to (\ref{46058534}).
For linear relations (\ref{47109986}) the energy density has the form
\begin{equation}
\label{40991956}
{\cE} = {\cEp}\doteqdot \frac{1}{8\pi}\left(\bDem^2 + \bBem^2\right)
=\frac{1}{8\pi}\left(\bDB\cdot\hconj{\bDB}\right)
\;\;,
\end{equation}
where $\hconj{\bDB}$ is complex conjugation to $\bDB$.

The components for quasitensor of angular momentum density are
\begin{equation}
\label{69808677}
\AMT^{\mu\nu\rho} \doteqdot x^{\mu}\,\EMT^{\nu\rho} -x^{\nu}\,\EMT^{\mu\rho}
\;\;.
\end{equation}

Consider a three-dimensional volume $\Vols$ bounded by two-dimensional closed surface ${\Surc}$
with directed element ${\dSurcb}$.
Let the volume and surface is independed of time for used coordinate system.
The energy-momentum of field in the volume $\Vols$ is
\begin{equation}
\label{70658589}
{\Energy}_{\Vols} \doteqdot {\EMV}^{0}_{\Vols}\doteqdot\int\limits_{\Vols}{\cE}\,\dVols
\;\;,\quad
{\EMV}^{i}_{\Vols}\doteqdot \int\limits_{\Vols}{\cP}^{i}\,\dVols
\;\;.
\end{equation}
Define similarly the antisymmetrical tensor of angular momentum
\begin{equation}
\label{71094566}
{\AMB}^{\mu\nu}_{\Vols}\doteqdot \int\limits_{\Vols}{\AMT}^{\mu\nu 0}\,\dVols
\;\;.
\end{equation}
For the components of ${\AMB}^{\mu\nu}_{\Vols}$ with zero index we have,
 according to definitions (\ref{69808677}) -- (\ref{71094566}),
 the following identity:
\begin{equation}
\label{68705015}
{\AMB}^{0 i}_{\Vols} \equiv x^{0}\,{\EMV}^{i}_{\Vols} - \xc^i_{\Vols}\,{\Energy}_{\Vols}
\;\;.
\end{equation}
Here the energy center coordinates $\xc^i_{\Vols}$ for the field in the volume $\Vols$ are introduced:
\begin{equation}
\label{69122437}
\xc^i_{\Vols} \doteqdot \frac{1}{\Energy_{\Vols}}\,\int\limits_{\Vols}x^{i}\,{\cE}\,\dVols
\;\;.
\end{equation}

In general case ${\EMV}^{\mu}_{\Vols}$, ${\AMB}^{\mu\nu}_{\Vols}$, $\xc_{\Vols}$ can be depend on time.

Let us designate the angular momentum vector with the same letter that the tensor:
\begin{equation}
\label{48715543}
{\AMB}_{\Vols i}\doteqdot \frac{1}{2}\,\epsilon_{ijk}\,{\AMB}^{ik}_{\Vols}
\;\;,\quad
{\AMB}_{\Vols}\doteqdot \left|{\bAMB}_{\Vols}\right|
\;\;,\quad
{\bAMB}_{\Vols} = \int\limits_{\Vols}{\mathbf{r}}\times{\bcP}\,\dVols
\;\;,
\end{equation}
where $\epsilon_{ijk}$ are the components of three-dimensional fully antisymmetrical unit tensor.

At last we have the following integral conservation laws for the field in the volume $\Vols$:
\begin{equation}
\label{70805791}
\frac{\df {\Energy}_{\Vols}}{\df x^{0}} = -\int\limits_{{\Surc}} {\EMT}^{0i}\,{\dSurc}_{i}
\;\;,\quad
\frac{\df {\EMV}^{i}_{\Vols}}{\df x^{0}} = -\int\limits_{{\Surc}} {\EMT}^{ij}\,{\dSurc}_{j}
\;\;,
%\end{equation}
%\begin{equation}
%\label{62239360}
\quad
\frac{\df {\AMB}^{\mu\nu}_{\Vols}}{\df x^{0}} = -\int\limits_{{\Surc}} {\AMT}^{\mu\nu p}\,{\dSurc}_{p}
\;\;.
\end{equation}

Consider some spatially localized solution of the model equations which corresponds to an electromagnetic particle.
This solution will be called particle one.
Let us designate the appropriate field as $(\partsol{\bDem},\partsol{\bBem})$.

Also let us define energy-momentum $\partsol{\EMV}^{\mu}$ of the the particle solution and
localization region $\partsol{\Vols}$ (bounded by $\partsol{\Surc}$) for the energy-momentum by the following way:
\begin{equation}
\label{44554599}
\partsol{\EMV}^{\mu} \doteqdot
\partsol{\EMV}^{\mu}_{\partsol{\Vols}}
\approx\mathop{\lim }\limits_{V \to \infty } \partsol{\EMV}^{\mu}_{\Vols}
\quad \Longleftrightarrow\quad
\forall \varepsilon  > 0\,,\,\,\,\,\,\,\exists\,\partsol{\Vols}:\;
|\partsol{\EMV}^{\mu}_{\partsol{\Vols}}
- \mathop{\lim }\limits_{V \to \infty } \partsol{\EMV}^{\mu}_{\Vols}| < \varepsilon
\;\;.
\end{equation}
Here $\varepsilon$ gives an accuracy for the approximate equation.
The definition of localization region is considered here with respect to some small time slice,
because $\partsol{\Vols}$ and $\partsol{\Surc}$ are considered to be constant.

Practically, the existence of the localization region of the energy-momentum is equivalent to existence
of finite limit $|\mathop{\lim }\limits_{V \to \infty } \partsol{\EMV}^{\mu}_{\Vols}|<\infty$.

Here we consider particle solutions for which
\begin{equation}
\label{51544218}
\Bigl|{\metr}_{ij}\,\partsol{{\EMV}}^{i}\,\partsol{{\EMV}}^{j}\Bigr| < \partsol{\Energy}^2
\;\;.
\end{equation}
For these solutions there is an inertial coordinate system in which $\partsol{\EMV}^{i}=0$.
These solutions correspond to massive particles. Their speeds are less than the speed of light.

Certain particle solution corresponds to a specified particle.
In practice we have many-particle configurations which must agree with some many-particle solution of the field model.
Here we consider the case when the particles of such solution are sufficiently distant from each other.
In this case the sum of appropriate particle solutions with time-dependent free parameters can be considered
as an initial approximation to the many-particle solution \cite{Chernitskii1999}.
An interaction between the particles will appear, specifically, as a dependence of their energy-momentums on time.

It is natural to define the particle's proper coordinate system
%of the particle
for which the momentum of the particle is zero and its energy center (\ref{69122437}) is at origin of coordinates:
 $\partsol{\EMV}^{i}=0$ and $\partsol{\xc}^{i}\doteqdot\partsol{\xc}^{i}_{\partsol{\Vols}}=0$ all the time.
But in the case of interaction this system is not inertial.
Thus it is convenient to define also the particle's proper inertial coordinate system in which
$\partsol{\EMV}^{i}=0$ and $\partsol{\xc}^{i}=0$ for given time point.
Let all quantities with respect to this system be designated by point at their bottom:
\begin{equation}
\label{72814093}
\partsol{\xp{\EMV}}^{i} \doteqdot 0
\;\;,\quad \partsol{\xp{\xc}}^{i} \doteqdot 0
\;\;.
\end{equation}

We have from identity (\ref{68705015}) and definition (\ref{72814093}) that
$\partsol{\xp{\AMB}}^{0 i} \equiv 0$ for the proper inertial coordinate system.
But we can make the transform from this system to another inertial coordinate system by formulas
\begin{equation}
\label{74194656}
\partsol{\xc}^{i} = \frac{v^{i}\, (\xp{x}^{0} - \xp{a}^0)}{\sqrt{1-v^2}}
\;\;,\quad
x^{0} = \frac{ (\xp{x}^{0} - \xp{a}^0)}{\sqrt{1-v^2}}
\;\;,\quad
\partsol{\EMV}^{i} = \frac{v^{i}\, \partsol{\xp{\Energy}}}{\sqrt{1-v^2}}
\;\;,\quad
\partsol{\Energy} = \frac{ \partsol{\xp{\Energy}}}{\sqrt{1-v^2}}
\;\;.
\end{equation}
Substituting of (\ref{74194656}) into (\ref{68705015}), we obtain the identity
\begin{equation}
\label{75419968}
 \partsol{\AMB}^{0 i} \equiv 0
%\;\;,
\end{equation}
for inertial coordinate systems obtained from the proper inertial one by time rotation
and time shift in the proper system (but without space shift).

A particle solution in proper inertial coordinate system may have a time dependent part.
Considering the time periodical dependence we can write:
\begin{equation}
\label{35335854}
\xp{\partsol{\bDB}} = \sum_{n=-\infty}^{\infty}\partsol{\xp{\bDB}_{\!|n}}(x^i)\,{\rm e}^{-{\bhim}\,n\,\omega\,x^{0}}
\;\;,
\quad
\xp{\partsol{\cE}} = \sum_{n=-\infty}^{\infty}\partsol{\xp{\cE}_{|n}}(x^i)\,{\rm e}^{-{\bhim}\,n\,\omega\,x^{0}}
\;\;.
\end{equation}

The mass of particle is its constant parameter. Thus we can use for the energy definition of mass
only the constant component of the energy density $\partsol{\xp{\cE}_{|0}}$.
But this component is determined also by time-dependent part of the solution.
Away from the localization region of the particle solution its field is small and we can write
according to (\ref{40991956}):
\begin{equation}
\label{44336412}
\partsol{\xp{\cE}_{|0}} = \frac{1}{8\pi}\,
\sum_{n=-\infty}^{\infty}\partsol{\xp{\bDB}_{\!|n}}\cdot\hconj{\partsol{\xp{\bDB}_{\!|n}}}
\;\;.
\end{equation}

It can be shown that the components with $n\neq 0$ in (\ref{44336412}) give
divergent (at infinity) energy integral. Thus for solutions with the non-vanishing
time dependent part (only in the representation at infinity)
it is not possible to introduce the localization region for energy-momentum, defined in (\ref{44554599}).
Here we investigate the case when the time dependent part at infinity is canceled:
$\xp{\partsol{\bDB}} = \xp{\partsol{\bDB}}_{|0}$.
Also we consider
$\partsol{\xp{\Energy}}=\partsol{\xp{\Energy}_{|0}}$, $\partsol{\xp{\bAMB}}=\partsol{\xp{\bAMB}}_{|0}$.
Taking into account these assumptions we define the mass and the spin by the following way:
\begin{equation}
\label{52468430}
{\mass}\doteqdot \partsol{\xp{\Energy}}
%(\partsol{\xp{\bDB}_{\!0}})
\;\;,\quad
{\bspin} \doteqdot \partsol{\xp{\bAMB}}
%(\partsol{\xp{\bDB}_{\!0}})
\;\;.
\end{equation}

For any inertial coordinate system we have the known expressions for energy-momentum:
\begin{equation}
\label{31059672}
\partsol{\Energy} = \frac{{\mass}}{\sqrt{1-v^2}}
\;\;,\quad
\partsol{\EMV}^{i} = \frac{{\mass}\,v^{i}}{\sqrt{1-v^2}}
\;\;,
\end{equation}
where $v^{i}$ are velocity components for the particle.

There are the static electromagnetic field configurations with spin.
This is, for example,
configuration with two point dyon singularities \cite{Chernitskii1999}.
Also a configuration with ring singularity \cite{Burinskii1974e} is worthy of attention
from this point of view.

On the border of the localization region the particle solution field is small
such that the constitutive relations can be considered as linear (\ref{65158987}).
The appropriate solution is known (see, for example, \cite{Chernitskii2005a})).
For the static part we have
\begin{equation}
\label{62150694}
\partsol{\xp{\bDB}_{\!|0}} = \C\,\frac{\mathbf{r}}{r^{3}} +
3\left({\bD}\cdot\mathbf{r}\right)\frac{\mathbf{r}}{r^5} - {\bD}\,\frac{1}{r^3}
+ \dots
\;\;,
\end{equation}
where $\C\doteqdot \Ce + \bhim\,\Cm$ is monopole moment or electromagnetic charge,
$\bD\doteqdot \bDe + \bhim\,\bDm$ is electromagnetic dipole vector,
$\Ce$, $\Cm$, $\bDe$, $\bDm$ are electric and magnetic charges and dipole moment vectors
for the particle solution.

Let us consider the interaction of particle with another particles which is sufficiently distant.
We investigate the initial approximation to appropriate many-particle solution in the considered particle
localization region. It has the form
\label{68895265}
\begin{equation}
\label{68902465}
\bDem = \partsol{\bDem} + \tilde{\bDem}
\;\;,
\quad
\bBem = \partsol{\bBem} + \tilde{\bBem}
\;\;,
\end{equation}
where $\tilde{\bDem}$, $\tilde{\bBem}$ are small field of the distant particle solutions,
the free parameters of space-time rotation and space shift for the solution $(\partsol{\bDem},\partsol{\bBem})$
are considered to be time-dependent.

For simplicity the problem is considered in proper inertial coordinates of the particle.
Let us write the field of the distant particles inside the considered localization region
in the following form of two Taylor series terms:
\begin{equation}
\label{68904378}
\xp{\tilde{\Dem}}_i = \xp{\tilde{\Dem}}_{i|} + \xp{\tilde{\Dem}}_{i|j}\,\xp{x}_j
\;\;,
\quad
\xp{\tilde{\Bem}}_i = \xp{\tilde{\Bem}}_{i|} + \xp{\tilde{\Bem}}_{i|j}\,\xp{x}_j
\;\;.
\end{equation}
where $\xp{\tilde{\Dem}}_{i|}$, $\xp{\tilde{\Bem}}_{i|}$, $\xp{\tilde{\Dem}}_{i|j}$, $\xp{\tilde{\Bem}}_{i|j}$
are independent of spacial coordinates.

Substitute of (\ref{68902465}) for (\ref{70805791}).
We take into account (\ref{62150694}), (\ref{68904378}), and linear Maxwell equations for small field
$(\tilde{\bDem},\tilde{\bBem})$. As result we have
\begin{eqnarray}
\label{72215262}
\frac{\df \partsol{\xp{\pmb{\EMV}}}}{\df x^{0}} &=& \Ce\,\xp{\tilde{\bDem}} + \Cm\,\xp{\tilde{\bBem}}
+ (\xp{\bDe}\cdot\nabla)\,\xp{\tilde{\bDem}}  + (\xp{\bDm}\cdot\nabla)\,\xp{\tilde{\bBem}}
\;\;,\\
\label{72217745}
\frac{\df \partsol{\xp{\bAMB}}}{\df x^{0}} &=& \xp{\bDe}\times\xp{\tilde{\bDem}} + \xp{\bDm}\times\xp{\tilde{\bBem}}
\;\;.
\end{eqnarray}
These equations coincide with classical motion equations of charged massive particle with spin and dipole moment
in external field for the proper inertial coordinate system of the particle.

Here the classical motion equations appear naturally as consequence of the integral conservation laws for the field
model.
But just these equations, considered as phenomenological, need mass, spin, charge,
and dipole moment of the particle. Thus here we have the substantiation for introduction of these
characteristics. Also the known phenomenological relation between mass and energy become valid.
\providecommand{\href}[2]{#2}

\end{document}